\documentclass[11pt,a4paper]{article}
\usepackage{jheppub}
\usepackage[utf8]{inputenc}
\usepackage[english]{babel}
\makeatletter
\usepackage{graphicx}
\usepackage{epstopdf}
\usepackage{mathtools}
\usepackage[strict]{changepage}
\usepackage[utf8]{inputenc}
\usepackage{physics}
\usepackage{hyperref}
\usepackage{mathrsfs}
\usepackage{amsthm}

\newtheorem{definition}{Definition}

\def\thesection{\arabic{section}}

\begin{document}

\title{Rényi mutual information inequalities from Rindler positivity}

\author[a]{David Blanco}
\author[b]{Leandro Lanosa}
\author[c]{Mauricio Leston}
\author[a]{Guillem P\'erez-Nadal}
\affiliation[a]{\it Departamento de F\'isica, FCEN - Universidad de Buenos Aires and Instituto de Física de Buenos Aires (IFIBA - CONICET), 1428 Buenos Aires, Argentina.}
\affiliation[b]{Departamento de Matem\'atica - FCEyN - Universidad de
Buenos Aires and IMAS - CONICET,  Pabell\'on I, Ciudad
Universitaria, C1428EHA Buenos Aires, Argentina.}
\affiliation[c]{\it Instituto de Astronom\'ia y F\'isica del Espacio (IAFE - CONICET), Universidad de Buenos Aires,
1428 Buenos Aires, Argentina.}

\emailAdd{dblanco@df.uba.ar, lanosalf@gmail.com, mauricio@iafe.uba.ar, guillem@df.uba.ar}

\date{\today}

\abstract{Rindler positivity is a property that holds in any relativistic Quantum Field Theory and implies an infinite set of inequalities involving the exponential of the Rényi mutual information $I_n(A_i,\bar{A}_j)$ between $A_i$ and $\bar{A}_j$, where $A_i$ is a spacelike region in the right Rindler wedge and $\bar{A}_j$ is the wedge reflection of $A_j$. We explore these inequalities in order to get local inequalities for $I_n(A,\bar{A})$ as a function of the distance between $A$ and its mirror region $\bar{A}$. We show that the assumption, based on the cluster property of the vacuum, that $I_n$ goes to zero when the distance goes to infinity, implies the more stringent and simple condition that $F_n\equiv{e}^{(n-1)I_n}$ should be a completely monotonic function of the distance, meaning that all the even (odd) derivatives are non-negative (non-positive). In the case of a CFT in 1+1 dimensions, we show that conformal invariance implies stronger conditions, including a sort of monotonicity of the Rényi mutual information for pairs of intervals. An application of these inequalities to obtain constraints for the OPE coefficients of the $4-$point function of certain twist operators is also discussed.}


\maketitle

\section{Introduction}\label{sect:1}
Entanglement entropy of a global state $\rho$ reduced to a spatial region $V$ is defined as the von Neumann entropy of the reduced density matrix $\rho_V$:
\begin{equation}
S=-\tr(\rho_V\log\rho_V)\,.
\end{equation}
This quantity is divergent in Quantum Field Theory (QFT), but it contains universal information that can be extracted from it. For instance, mutual information between two non-intersecting regions $A$ and $B$,
\begin{equation}
I(A,B)=S(A)+S(B)-S(AB)\,,\label{mutua}
\end{equation}
is a finite quantity that can be used to extract universal information from the entanglement entropy \cite{Casini:2009sr}. Mutual information is positive and increases upon adjoining a region $C$ to $B$, i.e.,
\begin{equation}
I(A,BC)\geq I(A,B)\,.\label{mono}
\end{equation}
This property is called monotonicity and it is equivalent to the strong subadditivity property of entanglement entropy
\begin{equation}
S(AB)+S(BC)\geq S(B)+S(ABC)\,.\label{ssa}
\end{equation}
Other interesting measures of entanglement are the Rényi entropies
\begin{equation}
S_{\alpha}(V)=\frac{1}{1-\alpha}\tr(\rho_V^\alpha)\,,
\end{equation}
where $\alpha \neq 1$ is a positive real number. The entanglement entropy $S(V)$ can be obtained from the Rényi entropies $S_{\alpha}(V)$ by taking the limit $\alpha\rightarrow 1$. Rényi mutual information (RMI) is just a generalization of equation (\ref{mutua}) for the Rényi entropies:
\begin{equation}
I_\alpha(A,B)=S_\alpha(A)+S_\alpha(B)-S_\alpha(AB)\,.
\end{equation}
Unlike entanglement entropy, Rényi entropies do not necessarily fulfill the strong subadditivity property in equation \cite{noSSA}. This tell us that mutual Rényi information is not necessarily a positive quantity and it does not satisfy the monotonicity property given by equation (\ref{mono}).

In this paper we show that a general property of relativistic QFT, known as {\it Rindler positivity} \cite{Casini:2010nn,Casini:2010bf}, imposes several constraints on the Rényi mutual information $I_n(A,\bar{A})$ as a function of the distance between $A$ and $\bar{A}$ when the global state $\rho$ is the vacuum of the QFT. Here, $A$ is a fixed-time region and $\bar{A}$ is the region obtained by making a reflection of one spatial coordinate.

Let us make a short summary of Rindler positivity: in \cite{Casini:2010nn,Casini:2010bf}, using Tomita-Takesaki theory, the following inequality was derived:
\begin{equation}
\left\langle 0\vert {\cal O}(\bar{A}){\cal O}(A)\vert 0\right>\geq 0\,,\label{Rindler}
\end{equation}

where ${\cal O}$ is an observable associated to a spacetime region $A$ in the right Rindler wedge ($\vert{x}\vert>{t}$), and $\bar{A}$ is the time and wedge reflection of $A$, i.e, the region obtained by making the transformation $(t,x,y,z)\rightarrow{-t,-x,y,z}$. For simplicity, we will restrict to the case in which $A$ is a region at fixed time $t=0$; in that case, we only need to make a reflection with respect to the spatial coordinate $x$. For a collection of $N$ spacetime regions $A_i$ equation (\ref{Rindler}) implies that the $N\times{N}$ matrix of coefficients \cite{Casini:2010nn}
\begin{equation}
M_{ij}=e^{(n-1)I_n(A_i,\bar{A}_j)} \,
\end{equation}
has to be positive definite for integer $n\neq 1$. This gives a set of inequalities coming from the fact that all the minors of the matrix $(M_{ij})$ have to be non-negative. These inequalities are non-linear expressions of the mutual information, with the exceptional case in which $N=2$, where we get the linear relation $I_n(A,\bar{A})+I_n(B,\bar{B})\geq{2I_n(A,\bar{B})}$. It is natural to ask what information can be extracted in general from all these expressions.

\begin{figure}[t]
  \centering
  \includegraphics[scale=0.7]{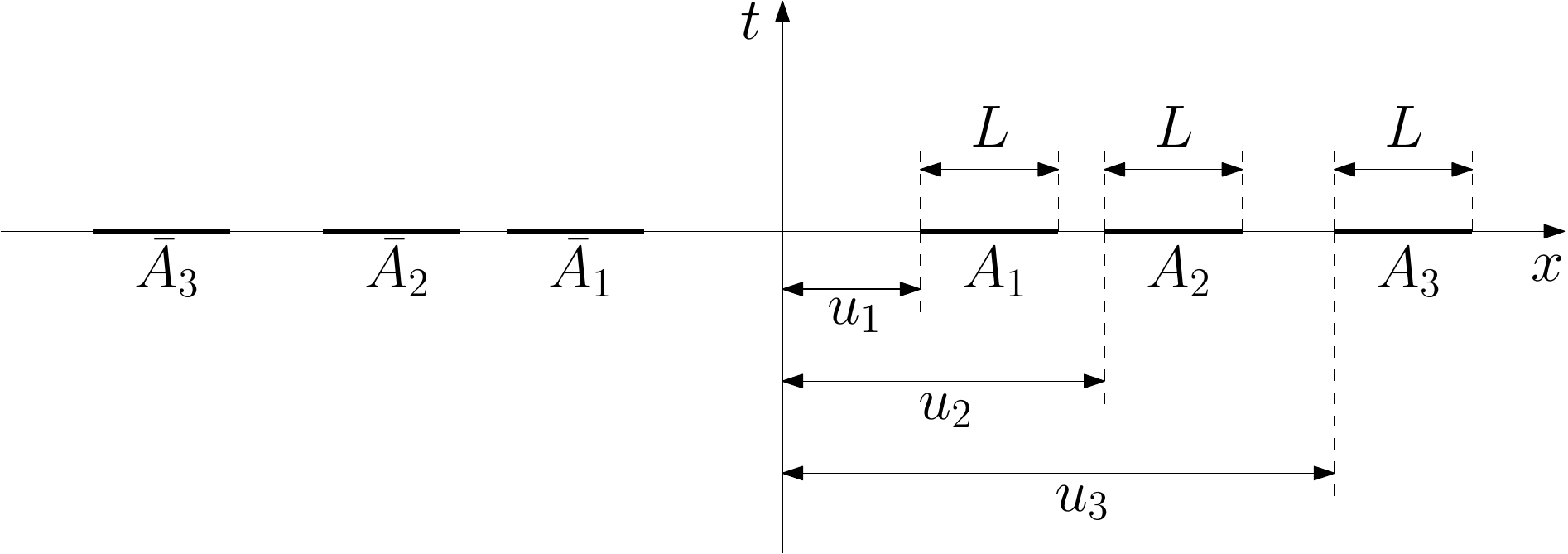}
  \caption{The family of regions considered consists of intervals $A_i$ that are obtained by making arbitrary positive translations of the fundamental interval $(0,L)$, and their wedge reflected counterparts $\bar{A}_i$. In the figure we see three different translations of the fundamental interval with parameters $u_1$, $u_2$ and $u_3$. The distance between the region $A_i$ and $\bar{A}_j$ is just $u_i+u_j$.}\label{intervalos}
\end{figure}

The situation which will allow us to go further with the implications of the inequalities is the following (for the sake of simplicity we momentarily think of the $1+1$ case). Consider the family of regions $A_i$ in the positive semi axis $x$ obtained by applying arbitrary translations of distance $u_i$ to a single region of a fixed length $L$ with left extreme point in the origin, see Figure \ref{intervalos}. Due to translation invariance, the RMI $I_n(A_i,\bar{A}_j)$ should be a function of the distance $\eta=u_i-(-u_j)=u_i+u_j$ between $A_i$ and $\bar{A}_j$, where $u_i$ and $u_j$ are the distances from the origin to the beginning of $A_i$ and $A_j$ respectively. Then, the coefficients $M_{ij}$ are just values of a single real variable function. We want to see what information can be extracted about $I_n(A_i,\bar{A}_j)$ as a function of $\eta$ from the positivity of the matrix $(M_{ij})$ (see \cite{Casini:2012rn} for a related study with different purposes).

The characterization of a real function $f$ of a single (real) variable such that the matrix defined by $M_{ij}=f\left(\frac{x_i+x_j}{2}\right)$ (for $x_i$ in $(a,b)$, $i=1,\,...,\,N$, for all $N$) is positive definite has been studied before \cite{Bernstein,2016arXiv160804010J,Widder,Berg}. In this paper we show that, when some of these results are supplemented with the additional condition that $I_n$ goes to zero when the distance between the regions goes to infinity \footnote{This can be seen by writing $\tr(\rho^n)$ as a correlation function of twist operators \cite{Calabrese:2009qy} and using the clustering property of the vacuum. See also \cite{Cardy:2013nua}.}, more restrictive conditions emerge for $I_n$ as a function of $\eta$. Schematically, these new inequalities put bounds to the derivatives of order $N$ of $I_n$ in terms of the lower order derivatives.

\textbf{Organization of the paper.}
We start this manuscript with a brief revision of some theorems on positive definite functions in section \ref{sect:2}. These results combined with Rindler positivity allow us to derive a set of inequalities for general QFTs, that are presented in section \ref{sect:3}. In section \ref{sect:4} we study the case of Conformal Field Theories (CFTs). In subsection \ref{sect:4.1} we rewrite the inequalities in terms of a cross-ratio when the regions involved are intervals in $1+1-$CFTs and show that Rényi mutual information is monotonous, or equivalently, that Rényi entropy is strong subadditive. After that, we show in subsection \ref{sect:4.2} how conformal symmetry allows us to obtain more restrictive inequalities. The inequalities obtained are checked in some known examples in subsection \ref{sect:4.4} and we present an application of one of the relations derived in subsection \ref{sect:4.5}. In section \ref{sect:5} we briefly comment the inequalities for the Rényi entropy. Finally, in section \ref{sect:6}, we discuss the relation between the inequalities we obtained and infinite divisibility.

\section{A brief review of results about positive definite functions}\label{sect:2}

The characterization and properties of positive definite functions of a single variable have been intensely studied in the early twentieth century, mainly by Schoenberg, Widder and Bernstein. Several known properties of the vacuum correlation functions in relativistic QFT are obtained by applications of some of these results. In this section, we give some definitions and enunciate key theorems on positive definite functions that are relevant to our paper. For a complete study of these topics we refer the readers to \cite{Bernstein,Widder,Berg} (also see \cite{2016arXiv160804010J} for a brief account of the main theorems).

There are two notions of positive definiteness for a function of a single variable. The definition we use here is the following:

\begin{definition}\nonumber
A real function $f:(a,b)\rightarrow{\mathbb R}$ is positive definite (PD) if, for any natural number $N$ and for any choice of points $\{x_i\}$ ($i=1,\,...,\,N$) with $x_i\in(a,b)$, the matrix $M$ of coefficients $M_{ij}\equiv{f}(\frac{x_i+x_j}{2})$ is positive definite \footnote{The other notion of positivity of a function arises when considering $M_{ij}=f(\vert{x}_i-x_j\vert)$ instead.}.
\end{definition}

Positive definiteness in this sense turns out to be a very restrictive condition. A surprising consequence of this property is the following: if $f:(a,b)\rightarrow{\mathbb R}$ is positive definite and continuous in $(a,b)$, then it is $C^{\infty}(a,b)$ (even more, it is real analytic there \cite{Widder}).

Moreover, the derivatives $f^{n}(x)$ of order $n$ satisfy an infinite set of inequalities valid at any $x\in(a,b)$: the $N\times{N}$ matrices ${H}^{(N,f)}$ of coefficients $\left({H^{(N,f)}}\right)_{m,n}=f^{(n+m)}$ ($n,m=0, ..., N-1$) are positive definite,
\begin{equation}
\det H^{(N,f)}=\left|
  \begin{array}{ccccc}
    f & f^{(0+1)} & f^{(0+2)}&.. &f^{(0+N-1)}  \\
    f^{(1+0)} & f^{(1+1)} & .&. & .. \\
    .. & .. & . & .& \\
    f^{(N-1+0)} & . & .&. & f^{(N-1+N-1)} \\
  \end{array}
\right|\geq{0}\label{det1}
\end{equation}

for all $N\in{\mathbb N}$. Conversely, an analytic function satisfying this infinite set of inequalities is PD. In fact, the inequalities (\ref{det1}) need only be satisfied at one point in $(a,b)$ and then they are automatically satisfied throughout the interval \cite{2016arXiv160804010J}.

An obvious consequence of the definition of positive definiteness is that a PD function is non-negative. A less obvious consequence is that the even derivatives of a PD function are also PD (this follows easily from the inequalities (\ref{det1})), and hence non-negative. Note also from the definition that a linear combination of PD functions with positive coefficients is also PD.

Simple examples of PD functions are $f(t)=e^{\lambda{t}}$ for $\lambda$ a real number. The positive definiteness can be checked easily both from the definition and from the inequalities (\ref{det1}). The definition of PD function requires that $\sum_{i,j=1..N}c_ic_jf(\frac{t_i+t_j}{2})\geq{0}$. In this case,  $\sum_{i,j=1..N}c_ic_jf(\frac{t_i+t_j}{2})=(\sum_{i=1...N} c_ie^{\frac{\lambda{t}_i}{2}})^2\geq{0}$. Therefore, $f$ is PD. Checking the inequalities is trivial since all the determinants are just $0$. Linear combinations of exponentials with positive coefficients will also be PD. In particular, a constant function $f(t)=c$, with $c\geq{0}$ is a PD function.

PD functions are closely related to {\emph{absolutely monotonic}} (AM) and {\emph{completely monotonic}} (CM) functions, whose definitions are the following:
\begin{definition}
A function $f$ is said to be absolutely monotonic (AM) if $f^{(n)}\ge 0$ for all $n=0,1,\dots$ and completely monotonic (CM) if $(-1)^nf^{(n)}\ge 0$ for all $n=0,1,\dots$.
\end{definition}
Note that the exponential $f(t)=e^{\lambda{t}}$ is AM for $\lambda>0$ and CM for $\lambda<0$. A PD function can always be written as the sum of an AM function and a CM function. This follows from a classical theorem, which states that a function $f$ on $(a,b)$ is PD if and only if it admits the following integral representation:
\begin{equation}\label{wieder}
f(t)=\int _{-\infty}^\infty e^{-\lambda{t}}g(\lambda)d\lambda=\int _{-\infty}^0 e^{-\lambda{t}}g(\lambda)d\lambda + {\int _{0}}^{\infty} e^{-\lambda{t}}g(\lambda)d\lambda
\end{equation}

where $g$ is non-negative (strictly speaking, $g(\lambda)d\lambda$ has to be understood as a Borel measure). Note that the first term on the right-hand side above is AM and the second term is CM.

Most important for this paper are the PD functions defined on $(0,+\infty)$ (or more generally on any interval of the form $(a,+\infty)$) which are bounded at infinity. From the decomposition (\ref{wieder}) it follows that such functions are necessarily CM. Roughly speaking, this is because the first term in (\ref{wieder}) diverges as $t\to\infty$, so this term must be absent in order for $f$ to be bounded at infinity (for a technical proof of this see \cite{Bernstein}). Conversely, it can be shown \cite{Widder} that any CM function on $(0,+\infty)$ admits the integral representation of the second term in (\ref{wieder}), and hence it is PD. In other words, {\it the space of PD functions on $(0,+\infty)$ which are bounded at infinity is equal to the space of CM functions on the same interval}.

This equivalence gives rise to additional inequalities to the ones given by equation (\ref{det1}), which come from the obvious fact that, if $f$ is CM, then $-f'$ is also CM. Using this and the above equivalence, we conclude that, for $f$ PD on $(0,+\infty)$ and bounded at infinity, $-f'$ is also PD.

The additional inequalities arise from substituting $f$ by $-f'$ in (\ref{det1}):

\begin{equation}
\det H^{(N,-f')}=(-1)^N\left|
  \begin{array}{ccccc}
    f' & f'^{(0+1)} & f'^{(0+2)}&.. &f'^{(0+N-1)}  \\
    f'^{(1+0)} & f'^{(1+1)} & .&. & .. \\
    .. & .. & . & .& \\
    f'^{(N-1+0)} & . & .&. & f'^{(N-1+N-1)} \\
  \end{array}
\right|\geq{0}\label{det2}
\end{equation}

Thus, PD functions on $(0,+\infty)$ which are bounded at infinity are characterized by two equivalent sets of conditions: (i) $(-1)f^{(n)}\geq{0}$ and (ii) equations (\ref{det1}) and (\ref{det2}). The first set of conditions appears to be much simpler than the second, but we will see that the second is more useful in some cases.

\section{Inequalities for relativistic QFT in $d+1$ dimensions}\label{sect:3}

\subsection{Implications of Rindler positivity for 2 and 4-point functions}

Before going to the case of the Rényi mutual information, let us first consider the implications of Rindler positivity for the two and four point functions of a real scalar field.
Let us consider a generic relativistic field theory corresponding to a scalar field $\phi$.  Let us define the state vector $\Psi=\sum_{i=1}^N c_i \phi(0,x_i,y,z)\Omega$ ($c_i$ real numbers for simplicity), where all the $x_i$ are positive, $x_i>0$, and $\Omega$ is the vacuum. The wedge reflected state $\bar{\Psi}$ is obtained by just replacing $x_i$ with $-x_i$. Rindler  positivity in this case asserts that $(\bar{\Psi},\Psi)\geq{0}$, which implies
\begin{equation}
    (\bar{\Psi},\Psi)=\sum_{i,j=1,..N}c_ic_j(\Omega,\phi(0,-x_j,y,z)\phi(0,x_i,y,z)\Omega)\geq{0}\,.
\end{equation}

Let us suppress in the notation the fixed value of the other coordinates.

Due to translation invariance, $(\Omega,\phi(-x_j)\phi(x_i)\Omega)$ will depend only on $x_i-(-x_j)=x_i+x_j$,
\begin{equation}
(\Omega,\phi(-x_j)\phi(x_i)\Omega)=f(x_i+x_j)\,.
\end{equation}

Rindler positivity implies that $f$ is PD. Moreover, using the cluster property (which implies that $f(x)$ goes to a constant when $x\rightarrow\infty$) we conclude that f is CM for any QFT.

A similar argument can be repeated for the 4-point function. A simple way to get a similar result for a function of 1 variable is to start from $\Psi=\sum_{i=1}^N c_i \phi(x_i)\phi(x_i+L)\Omega$, with a given positive $L$. In this way, using translation invariance, we will get that the function $f^{L}$ of one variable defined by

\begin{eqnarray}
    f^{L}(x_i+x_j)&=&(\Omega,\phi(-x_j-L)\phi(-x_j)\phi(x_i)\phi(x_i+L)\Omega)\nonumber\\
    &=&(\Omega,\phi(-L)\phi(0)\phi(x_i+x_j)\phi(x_i+x_j+L)\Omega)\,,
\end{eqnarray}
should be CM.

Let us notice that this simple constraint applies to a particular class of $4-$point functions $W(u_1,u_2,u_3,u_4)$, where $u_2-u_1=u_4-u_3\equiv{L}\geq{0}$. This and translation invariance tell us that $W$ is the function $f^L$ of the single parameter given by $u_3-u_2$. Let us remember that all the other coordinates are the same in each of the 4 points.

In order to show a use of these inequalities, let us show why $\frac{e^{-r^2/\lambda}}{r}$ cannot be a two point function for equal time points separated by a distance $r$ in any QFT. A simple check of the inequalities (\ref{det1}) and (\ref{det2}) shows that (\ref{det1}) with $N=2$ is violated. On the other hand, the two point function of a massive scalar field in $2+1$ dimensions,
$\frac{e^{-r/m}}{r}$, is the product of two CM functions, $e^{-r/m}$ and $1/r$, and hence it is CM as it should.

\subsection{Inequalities for the Rényi mutual information in $d+1$ dimensions}\label{sect:3.2}

In this section we explore the implications of Rindler positivity for the RMI between two regions. The main result of this section is a set of inequalities for the RMI between a spacelike region and its reflection that holds in any QFT for any dimensions.

Rindler positivity applies to any family of regions in the right Rindler wedge with all their reflections in the left Rindler wedge. But as we anticipated in section \ref{sect:1}, to use the results on positive definite functions of a single variable, the family of regions needs to be further restricted. A simple example of the construction of an allowed family is the one we illustrated before in figure \ref{intervalos}. More generally, the family of regions should fulfill the following two requirements:

\begin{enumerate}
\item The regions on the right Rindler wedge should be a one-parameter family of regions, in such a way that a real number $u_i$ fixes the region $A_i$.
\item The RMI of a pair $A_i\bar A_j$ should depend on $u_i$ and $u_j$ only via the sum $u_i+u_j$, $I_n(A_i\bar A_j)=I_n(u_i+u_j)$.
\end{enumerate}

\begin{figure}
  \centering
  \includegraphics[scale=0.8]{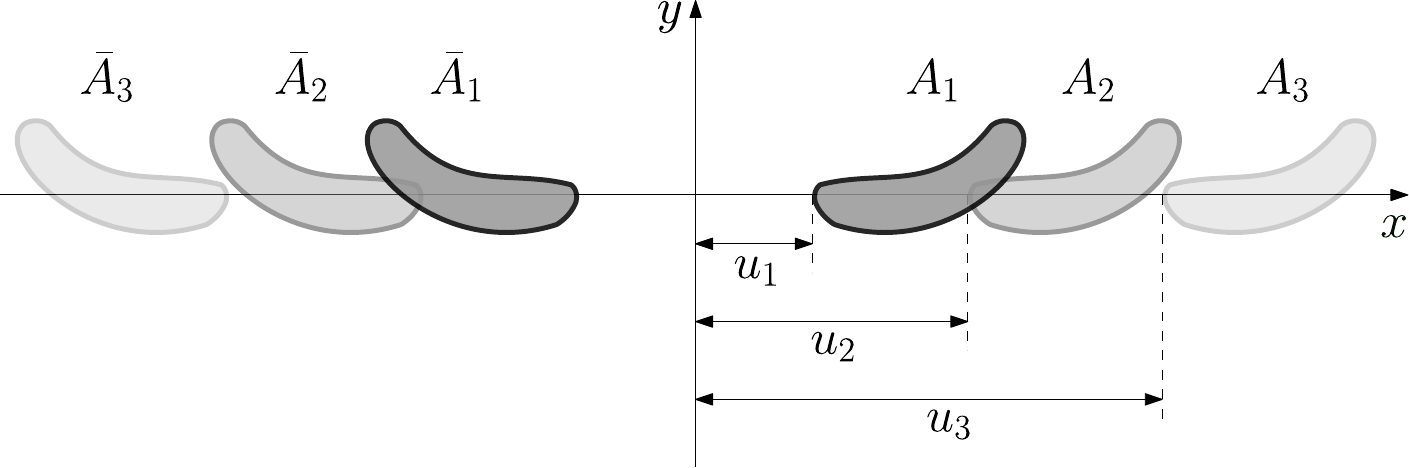}
  \caption{Some representatives of the family of regions located in the right Rindler wedge and their reflections. Each region in the right Rindler wedge is obtained by a translation in $x$ of the same region. The RMI between $A_i$ and $\bar{A_j}$ depends only on the distance $u_i+u_j$ between these regions.}\label{fig2}
\end{figure}

Note that these requirements constrain both the family of regions and the parameter used to label the regions. In figure \ref{fig2} we can see an example of a family that fulfills these requirements in $2+1$ dimensions. Each individual region in the figure is obtained by making an arbitrary positive translation in $x$ of the fundamental region located at $x=0$. In this case $I_n(A_i,\bar{A}_j)$ depends only on $u_i$ and $u_j$ via $\eta = {u_i+u_j}$, i.e, the sum of the distances to the origin of $A_i$ and $A_j$ respectively.

Rindler positivity applied to one of these families of regions implies that $F_n=e^{(n-1)I_n(\eta)}$ {\it is a PD function of $\eta$}. By construction, the distance $\eta$ covers the set $(0,+\infty)$; therefore, $F_n$ is a PD function in such interval. With the additional condition that $I_n$ goes to zero when the distance goes to infinity we can then conclude that {\it $F_n=e^{(n-1)I_n(\eta)}$ is a CM function} (see the discussion around equation (\ref{wieder})). As we explained at the end of section \ref{sect:2}, this implies that it satisfies the set of inequalities given by equations (\ref{det1}) and (\ref{det2}). We have thus derived a set of inequalities that the RMI between an arbitrary region and its reflection must satisfy \footnote{Note that the inequalities are written in terms of $I_n(A_i,\bar{A}_j)$, but since $A_j$ is a translation of $A_i$, $\bar{A}_j$ is the reflection of $A_i$ with respect to a plane located at the middle point between $A_i$ and $\bar{A}_j$.}.

Let us show some of the inequalities arising from equation (\ref{det1}) in the case $N=2$ and equation (\ref{det2}) for $N=1$. The first and simpler ones are the following

\begin{eqnarray}
\det\; H^{\left(2,f\right)}\geq{0}&\rightarrow&{I}_n''\geq{0}\label{ine1}\,,\\
\det H^{\left(1,-f'\right)}\geq{0}&\rightarrow&{I}_n'\leq{0}\label{ine2}\,.
\end{eqnarray}
The higher order inequalities are in general non-linear in the derivatives of $I_n$. For instance, the first following ones are ${I}_n^{(4)}I_n''+ 2(n-1)(I_{n}'')^3-{(I_n''')}^2\geq 0$ and ${(I_n)'(I_n)''' + (n-1)}I_n''(I_n')^2 - (I_n'')^2 \geq 0$.

In general, the first set of inequalities state that the highest order derivative appearing there (which is of order $2N$) will be greater than certain non-linear combinations of lower order derivatives. In the second set of inequalities, the highest derivative is $I^{2N-1}$ (odd) and the inequality also tells us that the $2N-1$ order derivative is bounded from above by certain expression that involves lower order derivatives.

Looking at equations (\ref{ine1}) and (\ref{ine2}), one could think that the alternating signs of these first derivatives are an indication that $I_n$ is a CM function. This does not follow from the previous inequalities, since the logarithm of a CM function is not a CM function and therefore $(n-1)I_n=\log(F_n)$ is not CM in principle. The inequalities obtained in general put lower and upper bounds for $I_n^{(2N)}$ and $I_n^{(N)}$  but they do not enforce $I_n$ to have alternating signs in their derivatives.

\section{Inequalities for CFTs}\label{sect:4}

In this section we focus on the special case of a conformally invariant QFT, where we study the obtained inequalities and we are also able to obtain more constraining relations using conformal symmetry. We verify the validity of the inequalities obtained for several concrete CFTs and we also show a simple application of the inequalities.

\subsection{Inequalities for intervals in a $1+1$ CFT}\label{sect:4.1}

In this subsection we show how to rewrite the inequalities obtained in section \ref{sect:3.2} in terms of a cross ratio, for the special case of intervals in $1+1$ dimensions. We will also show here that RMI exhibits a sort of monotonicity property when expressed in terms of the cross ratio.

Consider a $1+1$ CFT and the family of regions of figure \ref{intervalos}, that consists of fixed-length intervals. Due to conformal invariance, $I_n(A_i,\bar{A}_j)$ depends only on $\tilde{\eta}\equiv\frac{\eta}{L}$, where $\eta$ is the distance between $A_i$ and $\bar{A}_j$. This quantity can be expressed in terms of the usual cross ratio for the intervals $(u_i,v_i)$ and $(u_j,v_j)$
\begin{equation}
 x\equiv\frac{(v_i-u_i)(-u_j-(-v_j))}{(u_i-(-v_j))(v_i-(-u_j))}\,.
\label{crossratio}
\end{equation}
The relation between the cross ratio $x$ and $\tilde{\eta}$ is the following

\begin{equation}
\tilde{\eta}=\frac{u_i+u_j}{L}=\frac{1}{\sqrt{x}}-1\label{defeta}
\end{equation}

Note that, by conformal invariance, any pair of intervals (even if we allow intervals of different lengths) that have the same cross ratio $x$ will have the same mutual information. We can see how the inequalities obtained for $I_n(\eta)$ are rewritten when considering $I_n$ as a function of $x$ using equations (\ref{crossratio}) and (\ref{defeta}). For instance, inequality (\ref{ine2}) for the RMI in terms of the usual cross ratio $x$ is expressed as
\begin{eqnarray}
\centering{I_{n}'(x)\geq{0}}\label{restriccion}\,.
\end{eqnarray}
By simple algebraic manipulations we can see that the cross ratio $x$ associated with two disjoint intervals $A_i$ and $A_j$ increases when we replace $A_j$ by a larger region $\tilde{A_j}\supset A_j$. Therefore, we see that {\it $I_{n}'(x)\geq{0}$ expresses monotonicity of RMI, or equivalently, strong subadditivity of Rényi entropy}.

When rewriting equation (\ref{ine1}) in terms of the cross ratio we get
\begin{equation}
    2xI''_{n}(x)+3I'_n(x)\geq{0}\label{restriccion1}\,.
\end{equation}
This inequality puts a lower bound to the negativity of the second derivative, which in principle does not enforce RMI to be a convex function of $x$.

\subsection{A stronger inequality}\label{sect:4.2}

\begin{figure}[t]
  \centering
\includegraphics[scale=0.8]{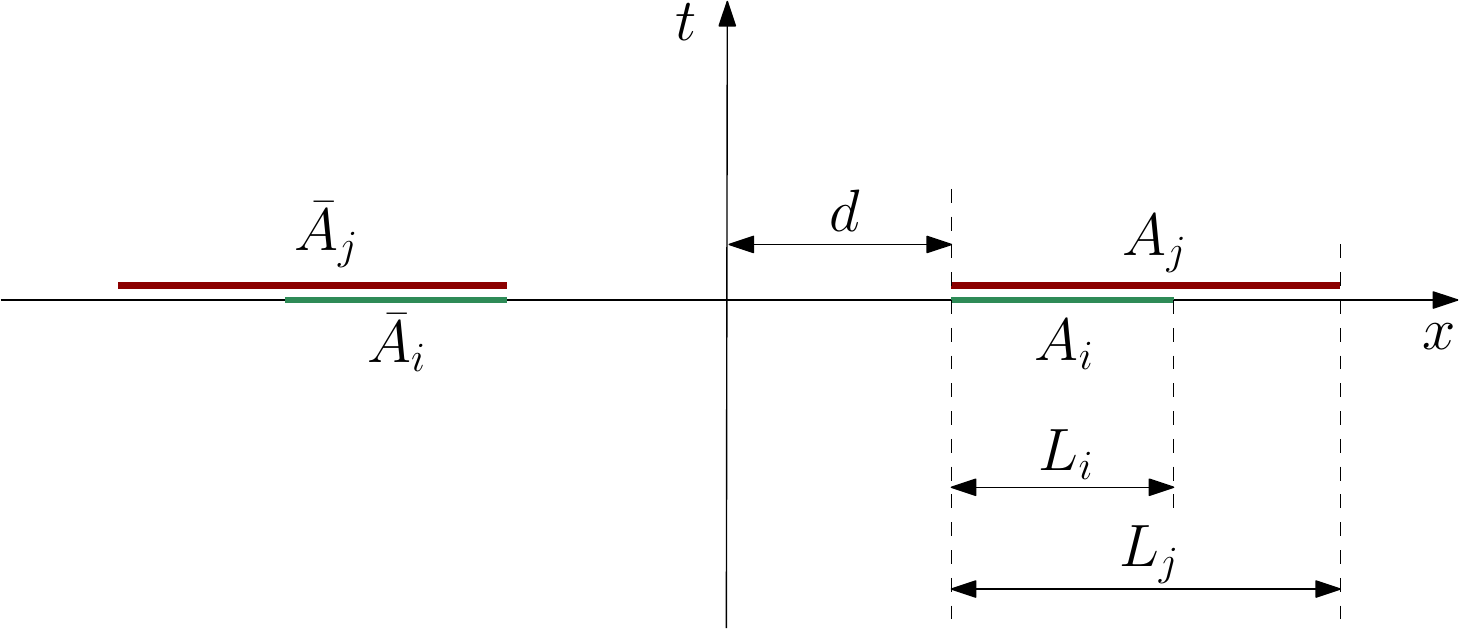}
  \caption{$A_i$ and $A_j$ are intervals of respective lengths $L_i$ and $L_j$ at $t=0$ with the same left starting point at $x=d$. The reflected regions $\bar{A}_i$ and $\bar{A}_j$ are also shown in the figure.}\label{fig3}
\end{figure}

Let us now consider the family of regions of figure \ref{fig3}. In contrast with the arrangement of figure \ref{intervalos}, in figure \ref{fig3} the family comprises all the intervals starting at the same point $x=d\geq 0$ (arbitrarily chosen) having different arbitrary lengths $L_i$, and their reflections.

The usual cross ratio $x$ associated to the pair $A_i,\bar{A_j}$ is
\begin{equation}
x=\frac{1}{(2d/L_i+1)(2d/L_j+1)}\,.
\end{equation}

Taking logarithms to both sides we get that
\begin{equation}
-\log(x)=\log(2d/L_i+1) + \log(2d/L_j+1)\,,
\end{equation}
is the sum of two arbitrary positive numbers covering the interval $(0,+\infty)$. Therefore, for this family, Rindler positivity tells us that $e^{(n-1)I_n}$ is CM as function of the new variable $\rho\equiv{-\log(x)}$ since (as we said before) we assumed that the exponential of $I_n$ is bounded when $\rho\rightarrow\infty$.

We want to remark that using conformal transformations, any pair of disjoint intervals with different lengths can be transformed into a pair of intervals with equal lengths. Therefore, for any pair of intervals both inequalities apply: those arising from complete monotonicity in $\rho$ as well as complete monotonicity as function of $\tilde{\eta}$.

Using results on compositions of CM functions, we can actually see that the complete monotonicity as a function of $\rho$ is stronger than the one as a function of $\tilde{\eta}$. $\tilde{\eta}$ is related to $\rho$ by $\rho=h(\tilde{\eta})=2\log(\tilde{\eta}+1)$. $h$ is a positive function whose first derivative is CM. It is easy to see that for such function, $f\circ{h}$ is CM if $f$ is CM, i.e., $I_n(\eta)$ will be CM if $I_n(\rho)$ is CM. The converse is not true, so complete monotonicity in $\rho$ will impose stronger conditions \footnote{CM in $\rho$ does not follow from CM of $f$ as function of $\tilde{\eta}$. It is enough to see a single example of a CM function $g$ in $(0,+\infty)$ such that $g\circ{h}$ is not CM. For instance, take $g(t)=e^{-\frac{1}{2}(t+1)}$ and check that $(g\circ{h})(\rho)$ has negative second derivative for $\rho<{2\log{2}}$.}. For instance, this condition enforces
\begin{equation}
xI_n''(x) + I_n'(x)\geq{0}\,,
\end{equation}
which is stronger than equation (\ref{restriccion1}). Writing these conditions in terms of $\tilde\eta$, we see that the stronger condition implies:

\begin{equation}
I_n''(\tilde\eta)\geq{-I}'_n(\tilde\eta)\frac{1}{\tilde\eta+1}
\end{equation}

This inequality can be written as $-(I'_n(\tilde\eta)(1+\tilde\eta))'\leq{0}$, saying that $-I'_n(\tilde\eta)$ should go as $\frac{1}{1+\tilde\eta}h(\tilde\eta)$ (for any distance), $h$ being a decreasing function of the distance.

The results of this subsection remain true in $d+1$ dimensions for the RMI of two arbitrary balls, because the latter depends only on the cross-ratio $x$ of the 4 points at which the boundaries of the balls intersect the line that joins their centers. That is, $F_n=e^{(n-1)I_n}$ is a CM function of $-\log(x)$ for an arbitrary pair of balls in a CFT in $d+1$ dimensions. 

\subsection{Check of the inequalities in some CFT models}\label{sect:4.4}

We have checked the set of inequalities $(-1)^nf^{(n)}\geq0$, $f=F_n=e^{(n-1)I_n}$ (where the derivatives are respect to $\rho=-\log(x)$ in the following $1+1$ CFTs: the massless free Dirac and scalar fields, the compactified free scalar and the critical Ising model. Let us next illustrate how simple the check of the inequalities turns out to be.

We start with the RMI between two intervals for the free fermion, which can be computed using the results for the Rényi entropy of an arbitrary set of intervals in \cite{Casini:2005rm}. In terms of the cross ratio $x$ given by equation (\ref{crossratio}), the RMI reads
\begin{equation}
I_n^{\textrm{free\;\;fermion}}(x)=-\frac{n+1}{12n}\log(1-x) \label{freefermion}
\end{equation}

This is an AM function since all its derivatives are positive. Therefore, the composition with $e^{-\rho}$ is automatically a CM funcion (since if $g$ is AM, and $h$ es CM then $g\circ{h}$ is CM - the converse is not true). Then, $I_n$ itself will be a CM function of $\rho$ implying that $F_n$ is a CM function.

In the case of the chiral scalar field \cite{Arias:2018tmw} the mutual information exhibit the same behaviour of being itself an AM function of $x$ (we checked this property numerically). In that case the check of the inequalities is then straightforward.

A perhaps more interesting situation is the case of the compactified free scalar, for which $I_2(x)$ is not an AM function (see equation 4.30 of \cite{Headrick:2010zt})

\begin{equation}
I_{2}(x) =\ln\left(\frac{\theta_3(il/R^2)\theta_3(ilR^2)}{\theta_3(il)\theta_4(il)}\right).
\end{equation}
where $R$ is the radius of compactification and $x$ is related to $l$ by $x=\displaystyle \frac{\theta_{2}^{4}(il)}{\theta_{3}^{4}(il)}$. Since the behaviour for small $x$ is $I_2\sim A\,x^{1/(R^2)}$, where $A$ is a positive number, for $R\neq{1}$ the second derivative becomes negative, while $I_2(x)$ and $I'_2(x)$ are positive. This means that $I_2(x)$ is not an AM function. Nevertheless, we checked that once we express $I_2$ as a function of $\rho$ it turns out to be a CM function and then the exponential of RMI fulfills the inequalities.

Another interesting model is the critical Ising model. For two disjoint blocks, the Rényi entropies have been computed in \cite{Alba:2009ek}. A closed expression for our function $F_2$ in that case is given by

{\small
\begin{eqnarray}
    F_{2}(x)&=&\frac{1}{(1-x)^{\frac{1}{8}}} \frac{1}{\sqrt{2}}\left\{\left[\frac{(1+\sqrt{x})(1+\sqrt{1-x})}{2}\right]^{1 / 2} +x^{1 / 4}+\left[(1-x) x\right]^{1 / 4}+(1-x)^{1 / 4} \right\}^{1 / 2}= \nonumber \\
    &=& \frac{1}{2(1-x)^{\frac{1}{8}}}\left[1+x^{\frac{1}{4}}+(1-x)^{\frac{1}{4}}\right]\,.
\end{eqnarray}
}

Replacing $x=e^{-\rho}$ in the last expression, it can be checked that $F_2$ is a CM function of $\rho$, though in this case $I_2$ itself is not CM.

RMI is also computed in holography and in those cases a phase transition occurs due to the large $c$ limit (see for example \cite{Headrick:2010zt}). The discontinuity in the derivative of RMI would immediately lead us to conclude that RMI is not a PD function in the holographic case. This is not a problem because holography involves taking a limit $c\rightarrow{\infty}$ and in this limit the PD character of $F_n$ may be lost.

\subsection{Constraining OPE coefficients with the inequalities}\label{sect:4.5}

In a $1+1$ CFT, the RMI between a pair of intervals can be written in terms of twist operators using the replica trick. It is known that, for two disjoint intervals, $e^{(n-1)I_n(x)}$ is a 4-point function of twist operators evaluated at $0,1,x,+\infty$, and as function of $x$, it admits the following convergent expansion (see for example equation 4.7 in \cite{Headrick:2010zt}):

\begin{equation}
e^{(n-1)I_n(x)}=x^{\frac{c}{6}(n-\frac{1}{n})}\left\langle\sigma_{1}(0) \sigma_{-1}(x) \sigma_{1}(1) \sigma_{-1}^{\prime}(\infty)\right\rangle=\sum_{m} c^{\sigma_{1}}_{ \sigma_{1} m }c^{m}_{ \sigma_{1} \sigma_{-1}} x^{d_{m}}\label{OPE}
\end{equation}

where the non-negative numbers $d_m$ are the conformal dimensions of the untwisted operators. For our discussion it is useful to collect all the terms with a given power $d$ of $x$,  writing this expansion as $\sum_{d} C_{d}x^{d}$. The expansion starts with the identity operator ($d=0$) with coefficient $1$ (in such a way that $I_n(0)=\log(1)=0$).

We want to see the constraints that CM as a function of $\rho$ imposes on the coefficients of the expansion. Replacing $x=e^{-\rho}$ we get
\begin{equation}
F_n=e^{-(n-1)I_n}=\sum_{d}C_d\;x^d=\sum_{d}C_{d} e^{-d\rho}\,.\label{expansion}
\end{equation}

One can show that $\sum_{i=1}^N c_ie^{-\lambda_it}$ (with $\lambda_i>0$) is CM if and only if every $c_i$ is non negative. Therefore, formally, the inequalities that we have derived imply that every $C_{d}$ should be non-negative. To best of our knowledge, this constraint on the coefficients $C_d$ had not been noticed before. Note that if all coefficients $C_d$ are positive, the above expression written in terms of $\tilde{\eta}$, $F_n=\sum_{d}C_{d} (1+\tilde{\eta})^{-2d}$, is also CM as it should.

In all the CFT models we studied the coefficients $C_{d}$ (obtained from adding up quadratic combinations of OPE coefficients) are actually non-negative. This can be seen by making a series expansion of $F_n$ in terms of $x$ and checking that each coefficient is in fact positive.

\section{Inequalities for Rényi entropy}\label{sect:5}

In \cite{Casini:2010nn,Casini:2010bf} Rindler positivity was applied directly to the exponential of the Rényi entropy $-S_n$ of two disjoint regions and then, by simple algebraic manipulations, it was shown that it also applies to the RMI. Our analysis implies that $g(d)=e^{-(n-1)S_n(d)}$ is a CM function of the distance $d$, between a pair of intervals of equal lengths (since $-S_n(A\,\bar{A})=I_n(A,\bar{A})-2S_n(A)$ and $S_n(A)$ is a constant). Notice that in this exponential appears the Rényi entropy of the union $A\bar{A}$.

\begin{figure}
  \centering
  \includegraphics[scale=0.8]{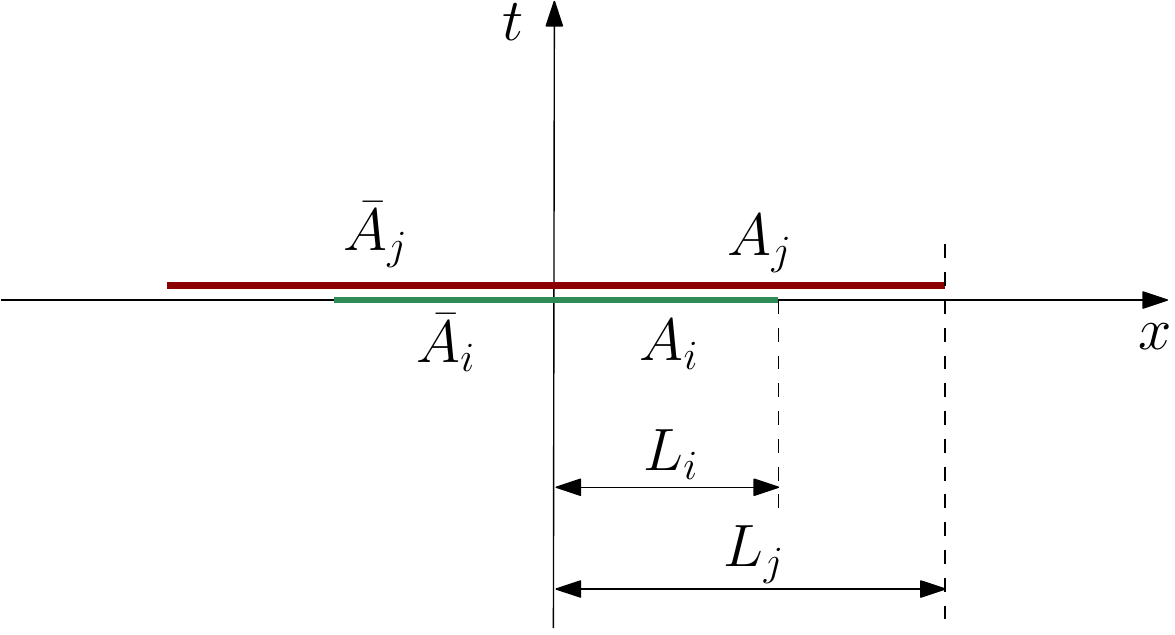}
  \caption{Two intervals $A_i$ and $A_j$ of different lengths starting at $x=0$ and their reflections. The union of $A_i$ with $\bar{A}_j$ is also an interval and its length is $L=L_i+L_j$}\label{Figure4}
\end{figure}

It would be interesting to derive inequalities for $S_n$ of a single region, considering $S_n$ as function of a parameter characterizing their size. In the case of intervals, for example, we can try to find inequalities for $S_n$ as a function of the length $L$. As usual, we need to choose a suitable family of regions. In this case, the relevant family is the one consisting of segments of different lengths, all of them starting at the origin (see figure \ref{Figure4}). In that case, $A_i\cup\bar{A_j}$ is again a single interval of length equal to $L_i+L_j$.

We cannot get complete monotonicity of $g(L)=e^{-(n-1)S_n(L)}$ as function of $L$ since we cannot impose that $S_n(L)$ goes to a finite value when $L$ goes to infinity. Therefore, we can only derive half of the inequalities, the ones given by equation (\ref{ine1}). Among other relations, we have that $S_n''(L)<{0}$. This comes just from positive definiteness of $g$.

\section{Remarks on some positivity conditions beyond Rindler positivity}\label{sect:6}

In the previous sections we obtained inequalities coming from a general theorem valid in any QFT (Rindler positivity). It is curious that, though these inequalities are not expected to hold for the entanglement entropy/mutual information, there are many cases in which the exponential of (minus) the entropy is a PD function. The standard example is the massless Dirac fermion in $1+1$ dimensions, in which $e^{-6 S}$ is in fact a correlator \cite{Casini:2010nn}.

Even in some cases where the function is not PD, it happens to be PD by pieces. For instance, we know that the phase transition appearing for the holographic mutual information between intervals in a $1+1$ assures that the exponential of $I$ cannot be PD since it is not an analytic function. However, it turns out that the exponential of $I$ is piecewise PD. Let us show this. The expression for the mutual information between two intervals is given by

\begin{equation}\label{I11}
I(x) =  \begin{cases} 0\,,\quad & x\leq 1/2 \\ (c/3)\ln(x/(1-x))\,,\quad & x\geq 1/2\end{cases}\,,
\end{equation}

Although this is not an AM function in the whole interval $(0,1)$ (since it is not differentiable at $x=1/2$), it is AM in $(0,\frac{1}{2})$ and $(\frac{1}{2},1)$ separately. Therefore, expressing $x$ in terms of $\rho$, the exponential of $\lambda{I}$ is

\begin{equation}\label{I12}
e^{\lambda{I(\rho)}} =  \begin{cases} \left(\frac{e^{-\rho}}{1-e^{-\rho}}\right)^{-\frac{c}{3}\lambda}\quad\quad\textrm{when} & \rho\leq \log(2) \\ 1\,\quad\quad\quad\quad\quad\quad\,\,\textrm{when} & \rho\geq{\log(2)} \end{cases}
\end{equation}
which is PD and CM by pieces, and this happens for any value of $\lambda$. This leads us to consider the issue of infinite divisibility.

\subsection{Infinite divisibility}
As we already noted before, our derivation of the CM character of $F_n=e^{(n-1)I_n}$ does not imply that $I_n$ itself is a CM function. If this was the case, then ${F_n}^{\alpha}=e^{\alpha(n-1)I_n}$ would be a PD function for every real positive $\alpha$. This last property is called infinite divisibility.

Infinite divisibility is not a consequence of Rindler positivity, but in many of the examples discussed in section \ref{sect:4.4} infinite divisibility occurs. In the cases where $I_n$ is itself an AM function of $x$ (as in the case of the free Dirac field in $1+1$, see equation (\ref{freefermion})), then it becomes a CM function of $\rho$ or $\eta$ after composition. In such cases, the exponential $e^{\alpha(n-1)I_n}$ is PD and CM for any $\alpha>0$, i.e, infinitely divisible.

In many examples, the first derivative of the entropy of a single interval is CM as a function of the length of the interval (see for instance \cite{Casini:2012rn}). This is exactly the necessary and sufficient condition for $e^{-\alpha{S}}$ to be a PD function for any $\alpha$ positive \footnote{
In \cite{Casini:2012rn} (in a different context, related to the implications of the conjectured condition by Fursaev of a path integral representation for the exponential of the entropy \cite{Fursaev:2006ih}) it was shown that infinite divisibility is equivalent to the condition that $-S''$ should be PD. Here, using that $S$ is positive, we arrive at a stronger and more manageable condition, which is that $S$ should be a Bernstein function, i.e. positive and with a CM first derivative.}, due to a well-known theorem \cite{Bernstein} which establishes that given a function $\psi:(0,+\infty)\rightarrow(0,+\infty)$, $f=e^{-\alpha{\psi}}$ is PD for any $\alpha >0$ if and only if $\psi '$ is CM. A function $\psi:(0,+\infty)\rightarrow(0,+\infty)$ with a CM first derivative is called a \textit{Bernstein function}. Bernstein functions can always be written as $a+b\,t-h(t)$, with $a$ and $b$ non-negative numbers and $h$ a CM function. For instance, the entropy of the vacuum reduced to an interval in a CFT is a Bernstein function. Bernstein functions grow at most linearly at infinity. Then, we have again infinite divisibility for the exponential $e^{-\lambda{S}}$.

\section{Summary and open questions}

In this paper, we studied how Rindler positivity together with the clustering property constrain the Rényi mutual information of certain pairs of regions. The inequalities we derived become more stringent when conformal symmetry is present. Let us summarize the main results obtained in this paper:

\begin{enumerate}
\item For general QFTs in arbitrary dimensions $e^{(n-1)I_n(A,\bar{A})}$ is a CM function of the distance $\eta$ between a region $A$ and its reflection $\bar{A}$ with respect to some plane. In particular, $I_n'(\eta)\leq{0}$ and $I''_n(\eta)\geq{0}$, that is, Rényi mutual information should be a decreasing convex function of $\eta$.

\item For CFTs in arbitrary dimensions, $e^{(n-1)I_n}$ is a CM function of $-\log(x)$ for an arbitrary pair of balls with cross-ratio $x$.

\item We verified that the inequalities derived are satisfied in many examples where an explicit expression for the Rényi mutual information is known.

\item We showed how the inequalities obtained can be used to impose non-trivial constraints on the coefficients appearing in the OPE of a 4-point function of certain twist operators.
\end{enumerate}

In this last spirit, we suspect that the inequalities derived here can be used to obtain more information about the structure of CFTs. We leave the study of this interesting topic for a future work.

As a final comment, we want to remark that the inequalities we derived come from Rindler positivity applied to particular families of regions (like the ones of figure \ref{intervalos} and figure \ref{fig3}). We have not explored yet the implications for more general regions, that will surely lead to stronger inequalities. In order to extract information of Rindler positivity for families depending on more than one parameter we will need results from positive definiteness of functions of several variables.

\section*{Acknowledgments}

The authors thank Nicol\'as Borda, Horacio Casini, Gaston Giribet and Juliana Osorio Morales for useful discussions. This work has been supported by UBA and CONICET.

\bibliography{Refs}

\end{document}